\documentclass[twocolumn,english]{revtex4-1}
\usepackage[T1]{fontenc}
\usepackage[latin9]{inputenc}
\usepackage{geometry}
\usepackage{graphicx}
\usepackage{esint}

\makeatletter
\@ifundefined{date}{}{\date{}}
\makeatother

\usepackage{babel}
\begin{document}
\title{Adaptable materials via retraining}
\author{Daniel Hexner}
\affiliation{Faculty of Mechanical Engineering, Technion, 320000 Haifa, Israel}
\begin{abstract}
Elastic metamaterials are often designed for a single permanent function.
We explore the possibility of altering a material's function repeatedly
through a self-organization, ``training'' process, controlled by
applied strains. We show that the elastic function can be altered
numerous times, though each new trained task imprints a memory. This
ultimately leads to material degradation through the gradual reduction
of the frequency gap in the density of states. We also show that retraining
adapts previously trained low energy modes to a new function. As a
result consecutive trained responses are realized similarly. We show
how retraining can be exploited to attain a response that would otherwise
be difficult. 
\end{abstract}
\maketitle
The ability to fabricate precise structures along with recent design
algorithms has allowed to construct metamaterials with new exotic
properties that are not commonly found in nature\citep{bertoldi2017flexible,chen2014nonlinear,coulais2016combinatorial,sussman2015algorithmic,boechler2011bifurcation,shelby2001experimental,kane2014topological,kushwaha1993,milton1992composite}.
Though successful, this results in mechanical devices with specialized
functions. Altering functionality after fabrication is usually impractical
since it necessitates to alter the structure. This is particularly
difficult if the degrees of freedom are on a microscopic scale and
buried in the bulk of a three dimensional material.

Nonetheless, altering a device's function could be highly advantageous,
allowing a ``general purpose'' material or device. That is, the
function can be adjusted for the task at hand, enabling a single programmable
device to replace multiple single function devices. This could potentially
be very important when weight of volume are very costly, as in the
case that they are launched into space\citep{shah2020shape,shah2021soft}.
We note that programmable materials perform a single function that
can be altered\citep{chen2021reprogrammable,silverberg2014using}
in contrast to multifunctional materials that are able to perform
simultaneously multiple functions\citep{lendlein2018multifunctional,rocks2019limits,torquato2002multifunctional}. 

In this letter we consider the viability of  reprogramming the function
of a material, using a recent approach of material training. Training
is based on the self-organization of the microstructure through plastic
deformations in response to external fields -- particularly, imposed
strains. By driving the system with carefully choreographed protocols,
the evolution of a material can be directed towards a desired elastic
response. Recent work has demonstrated that a generic disordered solid
can acquire highly complex non-linear functionality in both simulations
and experiments\citep{pashine2019directed,hexner2020periodic,hexner2020training,hexner2020effect}.
The advantage of self-organization, is that it does not necessitate
a direct manipulation of the micro-structure. Training can also be
thought of as a physical realization of a learning rule\citep{stern2020supervised,stern2020supervised2}.

We show that a material can be trained repeatedly, attaining each
time a different predefined elastic response. We first study how one
function evolves into another. Towards this goal we perform a normal
mode analysis. A strain response corresponds to a low frequency mode,
which is gapped from the remaining of the spectrum. Retraining, does
not introduce a new low frequency mode but rather adapts the previous
one. As a result there is a large overlap in how consecutive functions
are encoded in the system wide response. We show that this adaptability
can be exploited to achieve functions that are difficult to train.

We also show that each training task imprint a memory, which accumulates
with the number of trained functions. Ultimately, this leads to failure
to perform the trained task. This degradation is revealed in the density
of states, where as a material is retrained the frequency gap between
the lowest energy mode and the remaining of the spectrum slowly closes.
This implies that material with a larger gap are more robust. We also
find that degradation is accelerated by training at large strains.
The proposed scenario suggested retraining fails through the formation
of spurious low frequency modes that compete with the desired function.
\\
\emph{Model and training protocol: }We model an amorphous material
as a random network of springs that are derived from disordered packings
of repulsive soft spheres at zero temperature\citep{Ohern,Wyart2005Ann}.
The center of each sphere corresponds to a node and each overlapping
pair of spheres is attached with a spring of unit stiffness. Packings
are convenient ensemble since the coordination number, $Z=\frac{2N_{b}}{N}$
is easily tuned by varying the the imposed pressure \citep{Ohern}.
Here, $N_{b}$ is the number of bonds and $N$ is the number of nodes.
In the limit of zero pressure the networks are isostatic, and have
an anomalously long range elastic responses\citep{ellenbroek2006critical,lerner2014breakdown}. 

We consider bonds that evolve through plastic deformation in response
to the internal stresses. We assume that the rest length of the bond
evolves in proportion to the stress it experiences\citep{hexner2020periodic,hexner2020effect}: 

\begin{equation}
\partial_{t}\ell_{i,0}\propto k_{i}\left(\ell_{i}-\ell_{i,0}\right).\label{eq:lmodel}
\end{equation}
Here, $\ell_{i}$ is the bond length, $\ell_{i,0}$ is the rest length
and $k_{i}$ is the spring constant. This model can be considered
the Maxwell model for viscoelasticity\citep{maxwell1867iv} in the
limit where the elastic relaxation is fast with respect to the viscous
dynamics. 

\begin{figure}
\begin{centering}
\includegraphics[scale=0.5]{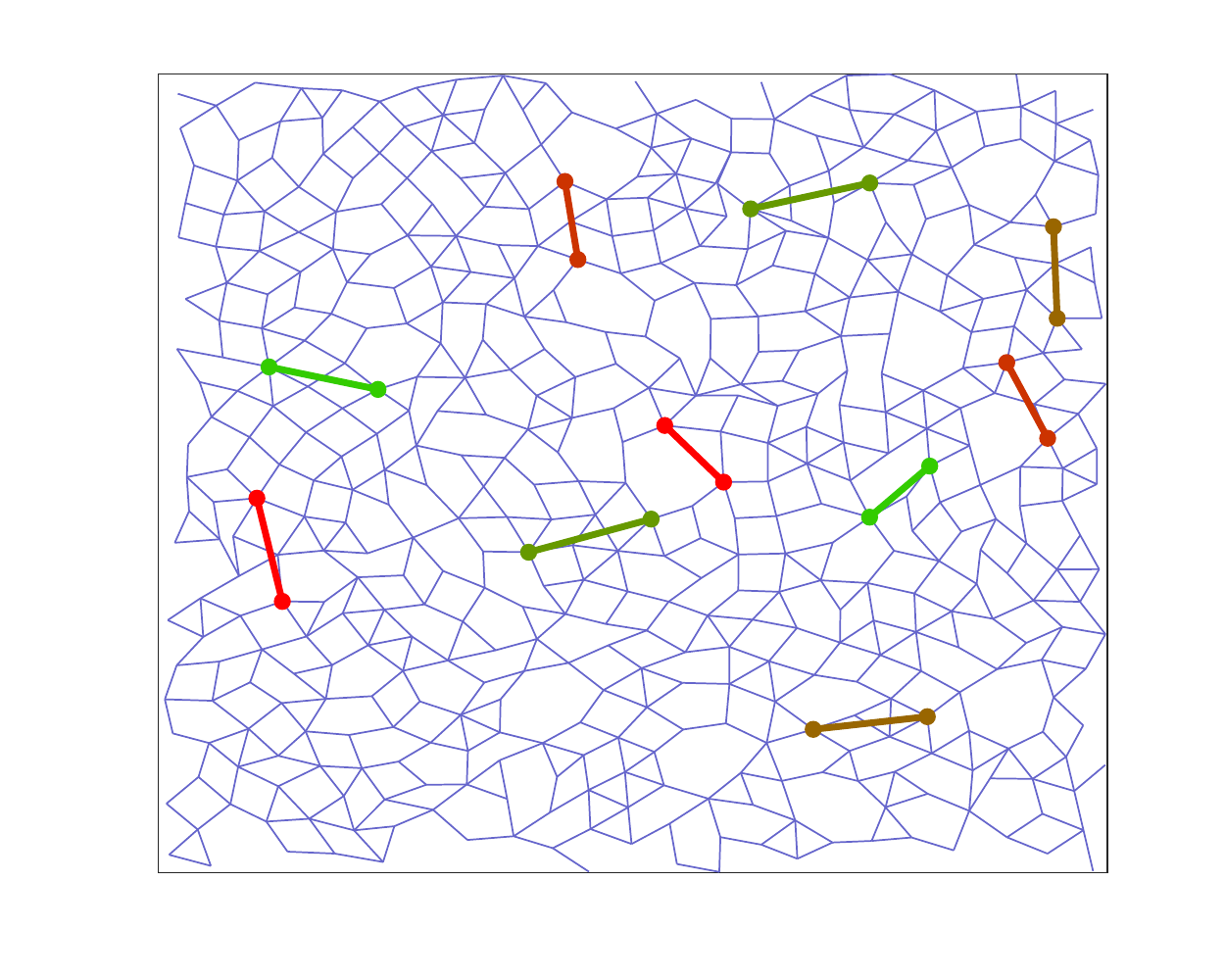}
\par\end{centering}
\caption{An example of a network with five pairs of source and target sites.
Each source and target are pairs of nearby nodes connected with a
line for clarity. We sequentially train pairs of source and target
(marked in the same color) to study how one function transforms into
another. Each function is trained with the aim of attaining a given
strain on the target in response to an applied strain on the source.
\label{fig:net}}
\end{figure}

We focus on training the simplest task -- the so called allostery
inspired response, where squeezing a ``source'' pair of nodes yields
a similar strain at a far away ``target'' site \citep{mitchell2016strain,rocks2017designing,yan2017architecture}.
Both the source and target are chosen to be at least half the length
of the system. We define the strain on a pair of nodes as the fractional
change in distance with respect to its unstrained value. Examples
of pairs of allostery sites are shown in Fig. \ref{fig:net}.

We follow the procedure of Ref. \citep{hexner2020periodic} to train
elastic responses. Both the source and target sites are repeatedly
strained quasistaticly up to an amplitude of $\epsilon_{Age}$ as
the system evolves through Eq. \ref{eq:lmodel}. This is implemented
numerically by discritizing time to small increments; at every time
step we vary the strain, minimize the energy to reach force balance
and then alter the rest lengths.

The intuition behind this strategy is that strain-response follows
low energy directions that couple the source and target, and therefore
the goal of training to create such a low energy ``valley''. Aging
at a constant strain decreases the energy at that imposed strain value.
This is seen by identifying the right hand side of Eq. \ref{eq:lmodel}
as the gradient of the energy with respect to $\ell_{i,0}$. Straining
periodically lowers the energy along the entire strain range, creating
a low energy mode that couples the source and target. 

Since we focus on training allosteric response that couples distant
sites, we resort to networks with a small deviation from isostaticity,
$\Delta Z\equiv Z-Z_{iso}\ll1$, characterized by anomalously long
range elasticity \citep{ellenbroek2006critical,lerner2014breakdown}.

\emph{Elastic response, and density of states:}\textbf{ }We sequentially
train task after task by straining periodically both the source and
target sites up to a strain of $\epsilon_{Age}$. Each task corresponds
to a different source and target sites both selected randomly. Our
goal is to train a system such that an imposed strain on the source
results in the same strain on the target. The convergence of the trained
response is characterized by straining the source up the training
strain $\epsilon_{Age}$ while measuring the strain on the target.
The error squared, $\left(\delta\epsilon\right)^{2}$, is defined
as the squared difference between the strain on the target and its
desired value, averaged over a measurement cycle.

\begin{figure}
\begin{centering}
\includegraphics[scale=0.7]{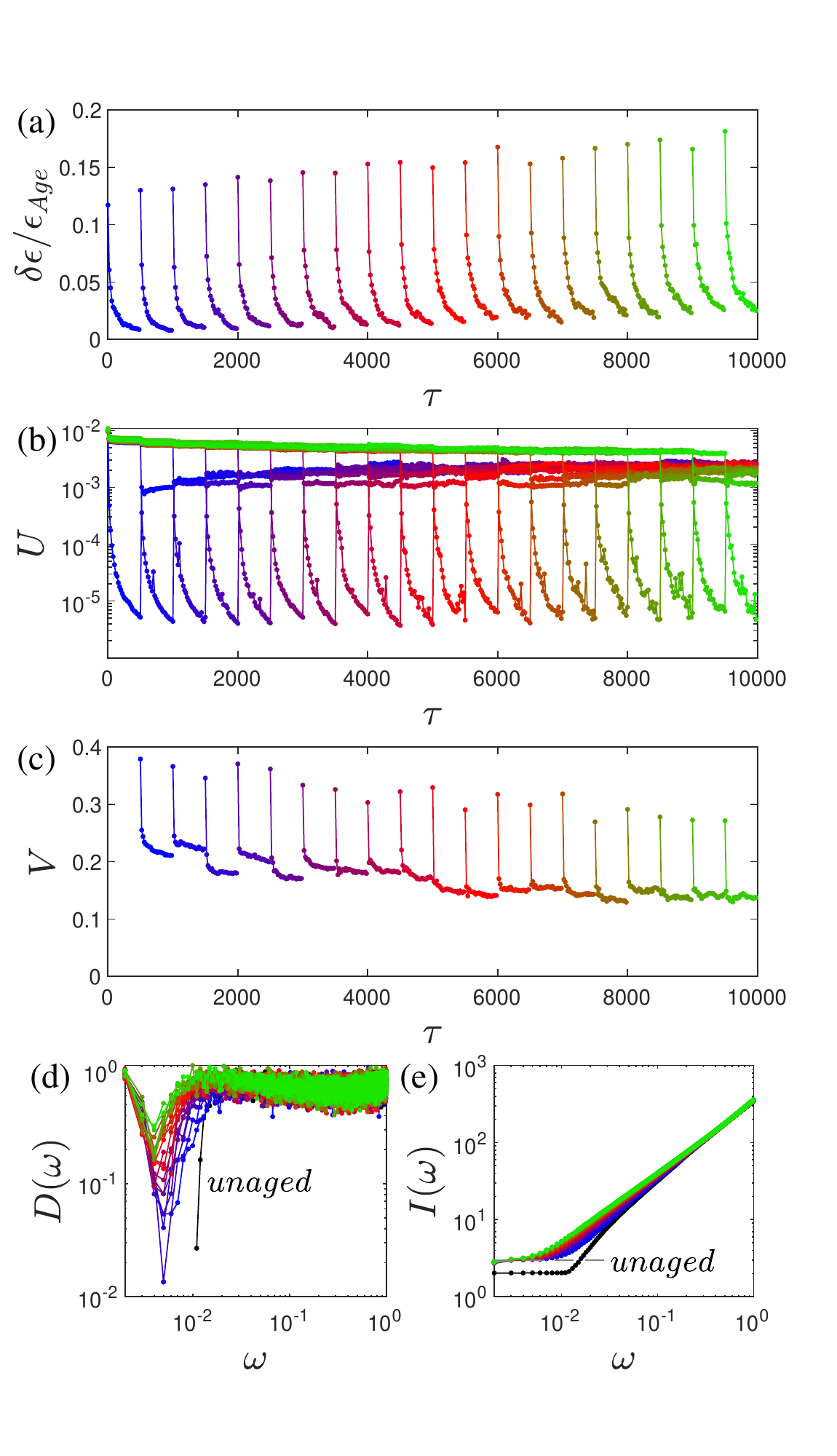}
\par\end{centering}
\caption{Sequentially training 20 allostery pairs, each denoted by a different
color. (a) The error in the trained response, slightly grows with
the number of tasks. (b) The energy for straining both the source
and target as a function of training cycles. Training reduces the
energy cost of the corresponding trained set. Note, the energy does
not return to its pretrained value when the task is switched. (c)
The overlap of the system-wide response between consecutive training
tasks. (d) The evolution of the density of states with number of trained
tasks. (e) The integrated density of states. Prior to training there
are the two zero modes associated with translations. Training adds
an third low energy mode (the dashed line marks $I\left(\omega\right)=3$).
Here, $N=512$, $\epsilon_{Age}=0.2$, $\Delta Z=0.038$. \label{fig:fig1}}
\end{figure}

Fig. \ref{fig:fig1}(a) shows $\delta\epsilon$ as a function of the
number training cycles for twenty different tasks, each depicted in
a different shade. Initially, the error for the newly assigned training
task is large, but the error decreases with training cycles, until
the task is switched. Overall, the system is able to adapt to a new
task multiple times, however, the error slightly grows with the number
of altered task.

As noted the response to an applied strain is associated with a low
energy valley. Fig. \ref{fig:fig1}(b) shows the energy measured along
each of the training paths, by straining both the source and target.
As expected, when a given task is trained the energy decreases. Interestingly,
even after the training task is changed the energy does not return
to its value prior to training, and remains substantially smaller.
As further tasks are trained that energy slowly increases.

These findings suggest that the system retains a memory of previously
trained tasks. To probe the fate of low energy modes we study the
density of states, $D\left(\omega\right)$. This characterizes the
high dimensional energy landscape within linear response. We first
compute Hessian, the matrix of second derivatives of the energy with
respect to the node locations, and then diagonalize to find the eigen-frequencies;
these are then binned to compute $D\left(\omega\right)$. In Fig.
\ref{fig:fig1}(d) , we show its evolution as the system is retrained.
The low frequency regime is of particular interest and shows a gap
which we discuss below. To better visualize the low frequency regime
we also compute the integrated density of states, $I\left(\omega\right)=N\int^{\omega}d\omega'D\left(\omega'\right)$,
which allows to enumerate the low energy modes. 

Fig. \ref{fig:fig1}(e), shows that prior to training $I\left(\omega\right)$
has two zero modes that are associated with the two independent translations
in two dimensions (due to the periodic boundary conditions). After
training the first task, another low energy mode is created that is
gapped from the remaining of the modes; namely, $I\left(\omega\ll1\right)=3$
. In our initial networks the gap scales as, $\Delta\omega\propto\Delta Z$
\citep{Ohern,Wyart2005_epl}. Even after retraining several times,
$I\left(\omega\ll1\right)=3$, implying that there always a single
extra low energy mode. This suggest that the system adapts the previously
trained mode to a new function, rather than forming an additional
mode. \\
\emph{Overlap between consecutive responses:} The adaptation of one
function into another suggest that two consecutive functions are realized
in a similar manner. We therefore define an overlap function which
compares the response of all the nodes between two consecutive tasks.
We define $\delta x_{i}^{\left(r\right)}$ the motion of node $i$
in response to straining the source and target sites. Here $r$ denotes
the task number defined by the $rth$ pair of source and target. The
overlap function is given by, 
\[
V^{\left(r\right)}\left(t\right)=C_{0}\left|\sum_{i}\delta x_{i}^{\left(r-1\right)}\left(t_{0}\right)\delta x_{i}^{\left(r\right)}\left(t\right)\right|.
\]
Here $t_{0}$ denotes the time at which the training task changed
from task $r-1$ to $r$, and $t>t_{0}$ is measured when task $r$
is trained. We normalize $V$ so that it is bounded by unity, $C_{0}^{-1}=\sqrt{\sum_{i}\delta x_{i}^{\left(r-1\right)}\delta x_{i}^{\left(r-1\right)}}\sqrt{\sum_{i}\delta x_{i}^{\left(r\right)}\delta x_{i}^{\left(r\right)}}$
. If the motion is uncorrelated then the overlap will vanish for large
systems as $1/\sqrt{N}$ ($N$ is the number of nodes).

\begin{figure}
\includegraphics[scale=0.6]{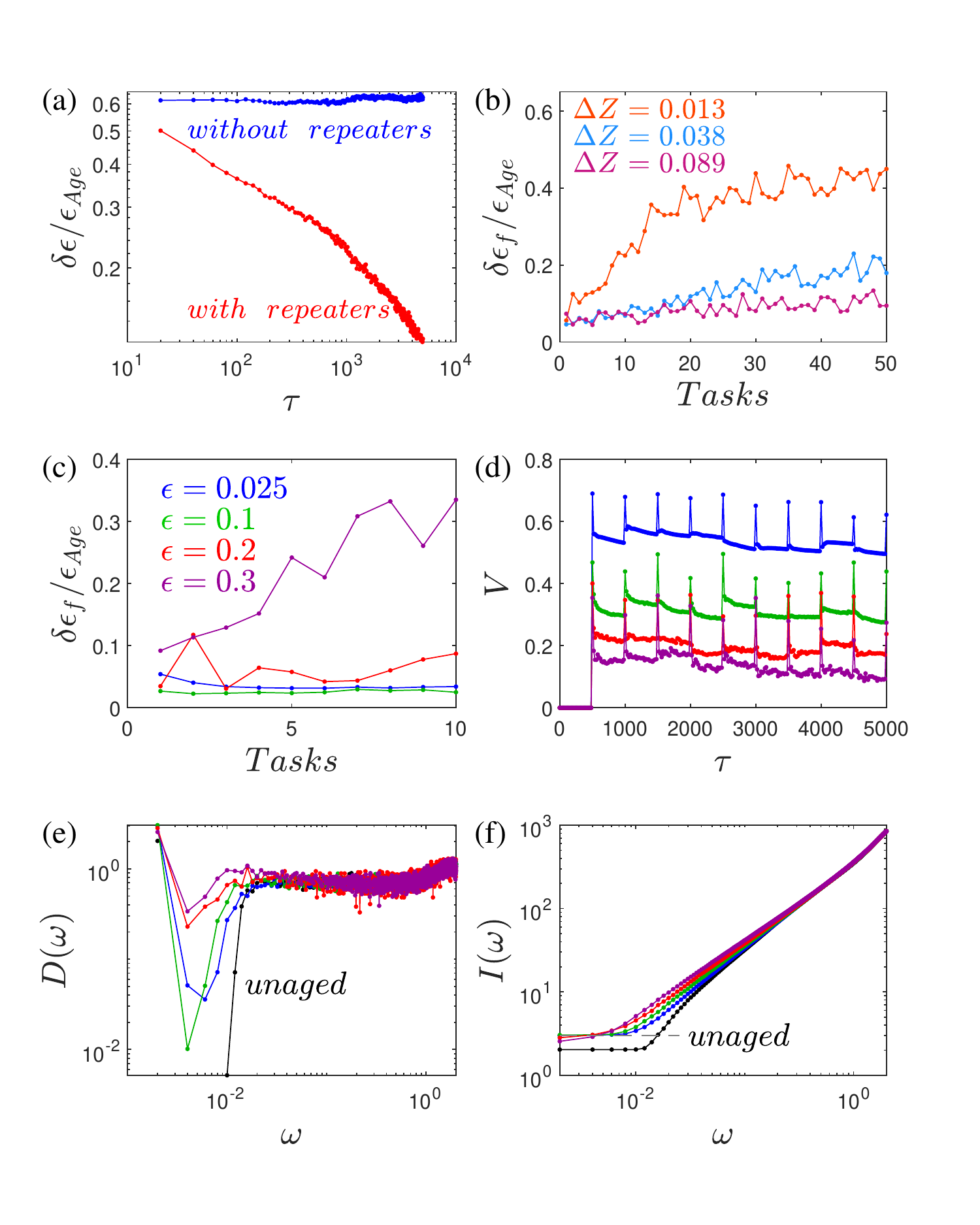}

\caption{(a) Training allostery in a highly coordinated network. The blue curve
shows that training fails when only the source and target are strained.
The red curve shows that training via ``repeaters'' an extended
low energy mode can be adapted by then straining only the source and
target. Here, $\Delta Z\approx0.52,\,\epsilon_{Age}=0.2$ and the
ratio of repeaters to number of nodes is $\approx0.2$. (b) Effect
of varying the coordination number: Networks with smaller $\Delta Z$,
and a smaller frequency gap, can be retrained fewer times. Here, $\epsilon_{Age}=0.2$
. (c-f) Effect of varying the strain amplitude. (c) Shows the error
after training is complete as a function of the number of tasks. At
large $\epsilon_{Age}$ the system degrades faster, leading the failure
of retraining. (d) The overlap function between consecutive tasks
grows as $\epsilon_{Age}$ decreases. (e) The density of states and
(f) the integrated density of states after ten tasks are trained.
Note that at large strains the gap is smaller. For (c-f) $\Delta Z\approx0.038.$
All networks have $N=512$ nodes. \label{fig:strain}}
\end{figure}

Fig. \ref{fig:fig1}(c) shows $V^{\left(r\right)}\left(t\right)$
as function of time for each of the training tasks. The overlap is
substantial even though different sets of source and targets are actuated.
That value does not decrease with system size, indicating that this
is not a finite size effect (see Supplemental material). We also consider
the effect of varying the training strain in Fig. \ref{fig:strain}
(d). We find that at small training strain the overlap is large. Training
at large strains necessitates larger changes to the structure \citep{hexner2020periodic}
yielding a smaller overlap. The finite overlap indicated a similarity
in the way that the system realizes both tasks, supporting the assertion
that one function transforms into another. 

We understand the large overlap as follows. Prior to retraining, there
is a single low energy mode corresponding to the function of the old
task. Elastic response follows the low energy modes and therefore
during training the new function, initially the motion couples strongly
to that mode. As a result, the new function shares a similarity between
the previous function. Similar behavior was found in Ref. \citep{husain2020physical}
in the context of epistasis.

\emph{Exploiting retraining to attain difficult responses:}\textbf{
}Up to now we have focused on training networks that are nearly isostatic,
whose long range elasticity was crucial. Straining only the source
and target do not couple distant sites in highly coordinated networks.
Ref. \citep{hexner2020periodic} introduced a strategy to access this
regime, that is based on additional ``repeater'' sites that are
strained in a similar manner to the source and target. These additional
sites are chosen randomly with a density that insures that repeater
sites are not too distant to couple. As a result the source and target
are able to couple through these intermediate repeaters. 

We consider a highly coordinated network, where an extended low energy
mode is trained using repeaters. We then alter the material's function
by retraining it by actuating only the source and target. Note that
the set of repeaters does not include the source and target sites.
Training via repeaters requires the actuation of a potentially large
number of sites, however these sites are non-specific. Namely, both
the sites are chosen randomly as well as the sign of the strain. We
believe, that this can be realized in experiment by having a small
fraction of the sites with an electric dipole moment; these are actuated
by applying a time varying electric field. Because the parameters
are chosen randomly the specific details of the driving is unimportant. 

The allosteric function is then trained by actuating only the specific
source and target sites. Fig. \ref{fig:strain}(a) compares the success
in training allostery with and without prior training with repeaters.
Adapting the old function trained with repeaters appears to converge
to the desired response, while in the system without prior training
the source and target do not couple. This demonstrates that adapting
an old function is far more successful than creating that same functionality
from scratch. 

\emph{Degradation:} Fig. \ref{fig:fig1}(d) shows that retraining
reduces the gap separating the mode associated with the function and
the remaining of the spectrum. Thus, the system softens as modes move
to lower energy. More importantly, the decrease of the frequency gap
affects the low energy spectrum which dictates the quasistatic response.
We argue, that this can be considered degradation since the additional
low energy modes compete with the trained function, leading to a larger
training error, as shown in Fig. \ref{fig:fig1}(a). Ultimately when
the gap closes we expect that the system fails to attain the trained
function. 

Such a scenario suggests that the gap plays an important role in the
ability to retrain the system. The gap prior to training depends on
the coordination number, $\Delta\omega\propto\Delta Z$ \citep{Ohern,Wyart2005_epl}.
We therefore expect that for small $\Delta Z$ , where the gap is
small, degradation occurs more quickly. Indeed Fig. \ref{fig:strain}(b)
shows that networks with a small $\Delta Z$ can be retrained fewer
times. Thus, there is a trade-off between large $\Delta Z$ where
the system can be retrained many times, and small $\Delta Z$ where
distant sites are easily coupled. 

\emph{Strain amplitude dependence:} We also consider the effect of
varying the training amplitude, $\epsilon_{Age}$. Fig. \ref{fig:strain}
(c) shows the response error at the end of each training set as a
function of the number of tasks. For the two smallest $\epsilon_{Age}$
degradation is barely visible after 10 tasks. At larger $\epsilon_{Age}$
the error grows with the number of trained tasks. This is consistent
with the diminished frequency gap in the density of states at large
$\epsilon_{Age}$, seen in Fig. \ref{fig:strain} (e) and (f). 

As noted, the overlap is large at small training strains, and decreases
with the strain amplitude (see Fig. \ref{fig:strain} (d)). 

\emph{Conclusions:} We have demonstrated that the function of a material
can be altered numerous time by retraining a disordered dashpot-spring
networks, enabling a material that can be programmed. Our approach
does not require the direct manipulation of the microstructure, but
is rather based on the autonomous evolution of the structure due to
the imposed strains. Our results suggest that retraining, repurposes
an old function rather than creating a new function. This, we argued,
is the result of the system following, during training, the already
present low energy mode of the previous trained function. Retraining
can be exploited to train functions that are otherwise difficult to
attain, such as, allostery in network with high connectivity. Training
a non-specific extended mode with repeaters, can then be retrained
to a specific allosteric response by just actuating the source and
target. This approach can potentially allow to couple distant sites
in experiments. 

While repurposing of low energy modes, has been beneficial, there
are cases where it is not. Particularly, if one would like to train
a number of different coexisting low energy modes. Our results suggest
this is challenging since training two set of different degrees of
freedom creates a single low energy mode. 

We have also shown that each trained function imprints a memory, which
ultimately leads to degradation. This is revealed by the normal mode
analysis, where retraining reduces the frequency gap between the low
energy mode associated with the function and the remaining of the
spectrum. We note that degradation is of a geometric origin since
the bond stiffnesses do not change. Ultimately the creation of competing
spurious low energy modes leads to failure. Systems with a larger
gap are more robust to retraining, while large strain amplitudes increase
the rate of degradation. We speculate that this could have relevance
to biological proteins that over the course of evolution alter their
function via mutations. Perhaps, the elastic behavior of a protein
encodes memories of their previous functionality. 
\begin{acknowledgments}
 I would like to thank Dov Levine, Andrea J. Liu, Sidney R. Nagel,
Nidhi Pashine and Menachem Stern for enlightening discussions. We
thank the ATLAS computing cluster for providing computer resources.
This work was supported by the Israel Science Foundation (grant 2385/20).
\end{acknowledgments}

\bibliographystyle{apsrev4-2}
\bibliography{biblo}

\end{document}